\documentclass[secnumarabic,amssymb, nobibnotes, aps, prd, twocolumn]{revtex4-2}
\usepackage{color}
\usepackage{psfrag}
\usepackage{dcolumn}
\usepackage{bm}
\usepackage{float}
\usepackage[latin1]{inputenc}
\usepackage[spanish,english]{babel}
\usepackage{amsfonts}
\usepackage{amssymb,epsf}
\usepackage{graphicx}
\usepackage{epstopdf}
\usepackage{amsmath,amssymb}
\usepackage{hyperref}

\begin{document}
	
		\title{\textbf Anisotropic Dark Matter Bosonic Stars in regularized 4D Einstein$-$Gauss$-$Bonnet gravity }
	\author{ Mohammad Mazhari$^{1,2}$}
	\email{m.mazhari@shirazu.ac.ir}
	\affiliation{$^{1}$ Physics Department and Biruni Observatory, Shiraz University, Shiraz 71454, Iran\\
		$^{2}$ Physics Department, Persian Gulf University, Bushehr 7516913817, Iran}

\begin{abstract}
		In this work, we have constructed anisotropic bosonic dark-matter star (DMS) solutions in the context of a regularized four-dimensional Einstein$-$Gauss$-$Bonnet (4D EGB) gravity theory. 
		Using dimensional regularization, we solve modified Tolman$-$Oppenheimer$-$Volkoff equations for a self-interacting complex scalar field in the dilute polytropic regime, $p_r = K \rho^2$, with anisotropy parameterized as $\sigma = \beta\, p_r \left( 1 - e^{-2\lambda} \right)$.
        We perform a comprehensive numerical analysis across the \((\alpha,\beta)\) parameter domain, where 
		\(\alpha \in [0,8]~\mathrm{km}^2\) and \(\beta \in [-2,0]\),
		to examine mass$-$radius relations and evaluate multiple stability indicators including static equilibrium \(dM/dp_c\), sound-speed causality, the radial adiabatic index \(\Gamma_r\), and energy conditions. 
		Positive Gauss$-$Bonnet coupling enhances both the maximum mass and compactness (e.g., \(M_{\rm max} \approx 1.62\, M_\odot\) at \(\alpha=0\) rising to \(\approx 2.09\, M_\odot\) at \(\alpha = 8~\mathrm{km}^2\)), while negative anisotropy reduces them (e.g., from \(\approx 2.21\, M_\odot\) at \(\beta=0\) to \(\approx 1.73\, M_\odot\) at \(\beta = -2\)). 
		The resulting configurations remain statically stable up to the mass peak and satisfy physical criteria. This work extends previous isotropic boson-star analyses by systematically incorporating anisotropy within a regularized 4D EGB framework. These findings provide observationally relevant predictions for compact dark-matter objects under modified gravity.
\end{abstract}
\maketitle

\section{Introduction}

Dark matter (DM) is still one of the unsolved problems in physics and cosmology. The dynamic evidence such as the rotation curves of galaxies and the precise measurements of the cosmic microwave background predict a considerable portion of the matter in the Universe to be non-luminous and to mostly interact via gravity \cite{Rubin1978,Rubin1979,Peterson1978}. The nature of DM at the microphysical level remains a mystery despite a lot of theoretical and experimental work. The most relevant and studied candidates are WIMPS \cite{Pospelov2008,Chang2014,Graesser2011}, and Bose-Einstein condensed scalar configurations \cite{Jungman1996,Boehmer2007,Sikivie2009,Harko2011,Chavanis2012,Harko2015}.


In the context of self-gravitating collections of bosons, bosonic (scalar-field) stars represent a well-studied relativistic framework considering the mass-radius relations and compact objects stability, suggested the stars serve as a theoretical laboratory for dark-matter phenomenology in high density regimes \cite{Colpi1986,Jetzer1992,Schunck2003}. Even in the absence of electromagnetic confinement, local pressure anisotropy (i.e. the difference between the radial $p_{r}$ and tangential $p_{t}$ pressures) occurs and has been shown to influence maximum mass, surface redshifts and stability criteria, and is present in dense stellar systems due to strong interactions, steep density gradients, and phase transitions in the matter \cite{Heintzmann1975,Mak2003,Doneva2012,Silva2015}.


Extensions of General Relativity focused on higher-curvature actions that include the Gauss$-$Bonnet (GB) term have started to modify the dynamics of gravity in the strong-curvature regimes around black holes and compact stars. The 4D EGB proposal suggests how to obtain nontrivial GB contributions in four dimensions via a dimensional regularization \cite{Glavan2020}. While the original construction prompted the discussion that the subsequent regularization/critique (see, e.g., \cite{Gurses2020,Hennigar2020}), has adequately addressed, several consistent regularized formulations have been useful effective four-dimensional descriptions in which the GB effects are nonvanishing.


In spite of the above developments, there has yet to be a dedicated, systematic work that integrates the microphysics of bosonic dark matter, macroscopic pressure anisotropy, and regularized 4D EGB corrections. This is important because (i) the interplay of anisotropy and GB-type corrections can alter the equilibrium sequences and stability thresholds \cite{Tangphati2021,Mazhari2024,Saavedra2025}, (ii) such changes may create detectable differences in the mass-radius relation, lensing configuration, or the signature of associated gravitational waves, and (iii) a cohesive treatment integrates underlying dark matter models (self-interacting scalars / BEC) and self-relativistic, phenomenological higher-curvature gravitational models \cite{Chavanis2012}.


This paper deals with the construction and analysis of equilibrium configurations of anisotropic bosonic dark-matter stars under a regularized 4D Einstein$-$Gauss$-$Bonnet framework. Specifically, we (1) derive the generalized hydrostatic equilibrium (modified TOV) equations corresponding to the regularized 4D EGB prescription \cite{Doneva2021,Gammon2024}, (2) construct a self-interacting bosonic equation of state driven by previous studies on boson stars, while incorporating a controlled anisotropy profile \cite{Boehmer2007,Lopes2024}, and (3) carry out numerical integrations and stability analysis (static stability checks, adiabatic index, and sound-speed causality) to understand and quantify how the GB and anisotropy parameters affect the mass-radius relations, density and pressure profiles, and overall physical viability  \cite{Dev2003}.

This paper is structured as follows: In Section 2, a summary of the field equations and the formalism of the regularized 4D EGB is presented. The equation of state of dark matter is formulated in Section 3. In Section 4, the stellar properties is subjected to numerical analysis is discussed. The stability analysis is causality and the energy conditions are discussed in Section 5. Section 6 concludes the paper and indicates possible future work.



\section{Review of dark matter bosonic star structure Equations in regularized 4D Einstein$-$Gauss$-$Bonnet gravity
}
To obtain the gravitational field equations corresponding to the gravitational theory described by the modified Einstein$-$Klein$-$Gordon action (or Einstein$-$Gauss$-$Bonnet$-$Klein$-$Gordon action), we start with the action in $D$$-$dimensions as described in Ref. \cite{Banerjee2021}:
\begin{eqnarray}
	S[g_{\mu \nu},\Phi] & = & \int d^4 x \sqrt{-g} \left( \frac{c^4 (R+\tilde{\alpha} \mathcal{G}_{GB})}{16 \pi G} + \mathcal{L}_M  \right),
	\label{action}
\end{eqnarray}
where $g$ is the determinant of the metric tensor, $\Phi$ is canonical complex scalar field, $\tilde{\alpha}$ is the Gauss$-$Bonnet coupling constant, which has dimensions of $[{\rm length}]^{2}$, and the Gauss$-$Bonnet invariant is defined by,
\begin{equation}
	\mathcal{G}_{\rm GB} \;=\; R^{\mu\nu\rho\sigma}R_{\mu\nu\rho\sigma} - 4 R^{\mu\nu}R_{\mu\nu} + R^2.
\end{equation}

To address the apparent singularity that arises when taking the limit as $D\to 4$, we define: 
\begin{equation}
	\tilde\alpha \equiv \frac{\alpha}{D-4},
	\label{eq:alpha_tilde}
\end{equation}
this rescaling ensures the theory avoids Ostrogradsky instabilities \cite{Motohashi2015}, enabling a well-defined novel formulation of 4D EGB gravity and then apply the finite regularized limit (refer to Refs.~\cite{Gammon2024,Banerjee2021}). Alternative dimensional-regularization or scalar$-$tensor embedding prescriptions might lead to inequivalent four-dimensional dynamics; accordingly, the prescription employed must be clearly stated. The four-dimensional regularized limit is employed in the present analysis.


Varying the action in Eq. (\ref{action}) with respect to the metric tensor $g_{\mu\nu}$ we obtain the field equations:
\begin{equation}
	G_{\mu \nu}+\frac{\alpha}{D-4}H_{\mu \nu}=\frac{8\pi G}{c^4} T_{\mu \nu}
	\label{eqgb4}
\end{equation}
where the Einstein tensor $G_{\mu\nu}$ and the Lanczos (Gauss$-$Bonnet) tensor $H_{\mu\nu}$ which are given by:
\begin{equation}
	G_{\mu \nu} \equiv R_{\mu \nu}-\frac{1}{2} R g_{\mu\nu}
	\label{Gmunu}
\end{equation}
\begin{eqnarray}\nonumber
	H_{\mu \nu} &\equiv& 2 \big(RR_{\mu \nu}-2R_{\mu \sigma} R^{\sigma}_{\nu}-2R_{\mu \sigma \nu \rho}R^{\sigma \rho}-R_{\mu \sigma \rho \delta}R^{\sigma \rho \delta}_{\nu} \big)\\&&-\frac{1}{2} g_{\mu \nu}\mathcal{G}_{GB}.
	\label{Hmunu}
\end{eqnarray}
These equations give the foundation for examining spherically symmetric stellar configurations in the regularized 4D EGB framework.

Let us proceed to derive the equations of hydrostatic equilibrium for a static, spherically symmetric spacetime which is represented by the metric~\cite{Mazhari2024,Banerjee2021}:
\begin{equation}
	ds^{2}=-e^{2\nu(r)}(cdt)^{2}+e^{2\lambda(r)}dr^{2}+r^{2}d\Omega^{2},
	\label{metric}
\end{equation}
Where the functions $\nu(r)$ and $\lambda(r)$ depend on the radial coordinate $r$, and $d\Omega^2$ is defined as
$
d\Omega^2 = d\theta^2 + \sin^2\!\theta\, d\phi^2
$.

Following Refs.~\cite{Moraes2021,Herrera2013}, an anisotropic fluid's energy$-$momentum tensor is expressed in this form:
\begin{equation}
	T_{\mu \nu}=({\epsilon} + {p}_{t}) u_{\mu} u_{\nu} + {p}_{t} g_{\mu\nu} - \sigma k_{\mu} k_{\nu},
\end{equation}

where ${\epsilon}={\rho}c^{2}$ is the energy density,   ${p}_{r}$ and ${p}_{t}$ denote the radial and tangential pressures, and $\sigma \equiv p_{t}-p_{r}$  represents the anisotropy factor. 
The four-velocity of the static fluid is $u^{\mu}=(e^{-\nu(r)},0,0,0)$, satisfying $u_{\mu}u^{\mu}=-1$. The unit radial vector $k^{\mu}$  obeys $k^{\mu}k_{\mu}=+1$ and $k^{\mu}u_{\mu}=0$.
\\

The covariant conservation of the energy$-$momentum tensor, $\triangledown_{\mu}T^{\mu}_{\nu}=0$, for 
$\nu=1$ leads to the hydrostatic equilibrium condition:
\begin{eqnarray}
	\frac{d{p}_r(r)}{dr}=-({\epsilon(r)}+{p}_r(r))\frac{d\nu(r)}{dr}+\frac{2{\sigma}}{r}.
	\label{stress}
\end{eqnarray}

To close the system, we define the mass function 
$m(r)$ via $e^{-2\lambda(r)}=1-\frac{2Gm(r)}{c^{2} r}$.
Substituting these relations into the field equations and using the above definition yields the modified Tolman$-$Oppenheimer$-$Volkoff (TOV) equations for 4D EGB gravity~\cite{Mazhari2024,Gammon2024,Banerjee2021}:
\begin{eqnarray}
	\frac{dp_{r}(r)}{dr}&=&-\frac{G\epsilon(r) m(r)}{c^\text{2} r^\text{2}}
	\big[\text{1}+\frac{p_{r}(r)}{\epsilon(r)}\big]\times
	\big[\text{1}-\frac{\text{2}Gm(r)}{c^\text{2} r}\big]^{-\text{1}}
	\nonumber\\
	&& 
	\times 
	\big[\text{1}+\frac{\text{4}\pi r^\text{3}p_{r}(r)}{c^\text{2}m(r)}-\frac{\text{2}G\alpha m(r)}{c^\text{2} r^\text{3}}\big]
	\nonumber\\&&
	\times
	\big[\text{1}+\frac{\text{4}G\alpha m(r)}{c^\text{2} r^\text{3}}\big]^{-\text{1}}
	+\frac{2 \sigma}{r} ,
	\label{pr1}
\end{eqnarray}
\begin{eqnarray}
	\frac{dm(r)}{dr}&=&\big[\frac{\text{4}\pi r^\text{2}\epsilon(r)}{c^\text{2}}+\frac{\text{6}\alpha Gm^\text{2}(r)}{c^\text{2}r^\text{4}}\big]\nonumber\\&&
	\times \big[\text{1}+\frac{\text{4}G\alpha m(r)}{r^\text{3}c^\text{2}}\big]^{-\text{1}}.
	\label{m1}
\end{eqnarray}

In the limiting case $(\alpha,\sigma)\to0$, the above equations recover the standard Tolman$-$Oppenheimer$-$Volkoff equation of General Relativity. For solving Eq. (\ref{pr1}) and Eq. (\ref{m1}), an appropriate equation of state must be chosen and the anisotropy factor must be determined. A full explanation of these two elements is given in the next section.

\section{Equations of State of an anisotropic bosonic dark-matter star}
\label{sec:eos}

Bosonic stars are self-gravitating configurations that can arise either from spin-zero fields, commonly called \emph{scalar bosonic stars}, or from spin-one fields, known as \emph{Proca stars}~\cite{Schunck2003,Brito2016,Herdeiro2020,Herdeiro2022}. 
The early studies concluded that, in the absence of self-interactions, the maximum mass of scalar bosonic stars is bounded by a specific upper limit~\cite{Kaup1969,Ruffini1969,Schunck2003}, 
and later researches indicated that self-interactions could significantly change this limit~\cite{Colpi1986,Seidel1991,Kusmartsev1991,Liebling2012,Cardoso2017}. 
Based upon the communication given in~\cite{Moraes2021}, we now present the EoS and anisotropy profile used in our investigation. In what follows, we describe the bosonic dark-matter model, its self-interaction potential, and the adopted anisotropy prescription.

We assume that dark-matter stars are described by a canonical complex scalar field, $\Phi$, minimally coupled to gravity and governed by the action introduced in Eq.(\ref{action}). In this framework, the matter Lagrangian of the dark-matter star is given by\cite{Liebling2012}:
\begin{equation}
	\mathcal{L}_M = - g^{\mu\nu} \partial_\mu \Phi \, \partial_\nu \Phi^* - V(|\Phi|),
	\label{eq:Lagrangian}
\end{equation}
where $V(|\Phi|)$ is the self-interaction potential. 
For static and spherically symmetric configurations we adopt the ansatz~\cite{Liebling2017}
\begin{equation}
	\Phi(r,t) = \phi(r)\, e^{-i\omega t},
	\label{eq:Phi_ansatz}
\end{equation}
where the real frequency $\omega$ characterizes the time-oscillatory phase of the scalar field.

The corresponding energy$-$momentum tensor is obtained from
\begin{equation}
	T_{\mu\nu}
	= -\frac{\partial \mathcal{L}_M}{\partial(\partial^\mu \Phi)} \, \partial_\nu \Phi
	-\frac{\partial \mathcal{L}_M}{\partial(\partial^\mu \Phi^*)} \, \partial_\nu \Phi^*
	+ g_{\mu\nu}\, \mathcal{L}_M .
	\label{eq:Tmunu}
\end{equation}

Although the scalar field itself depends on time, its stress$-$energy tensor remains time independent, and the 4D EGB gravitational field equations take the usual form corresponding to a fluid. In this case, the energy density, radial pressure, and tangential pressure are computed, respectively, as follows:
\begin{align}
	\rho &= \omega^2 e^{-2\nu} \phi^2 + e^{-2\lambda} \phi'^2 + V(\phi), \label{eq:rho}\\
	p_r &= \omega^2 e^{-2\nu} \phi^2 + e^{-2\lambda} \phi'^2 - V(\phi), \label{eq:pr}\\
	p_t &= \omega^2 e^{-2\nu} \phi^2 - e^{-2\lambda} \phi'^2 - V(\phi). \label{eq:pt}
\end{align}

Therefore, the pressure anisotropy, defined as $\sigma \equiv p_t - p_r$, is expressed as follows:
\begin{equation}
	\sigma = -2 e^{-2\lambda} \phi'^2 < 0 ,
	\label{eq:sigma}
\end{equation}
which represents a characteristic signature of boson stars~\cite{Chavanis2012}. 
Although boson stars are intrinsically anisotropic, under certain limits the anisotropy may be neglected and the system approximated as isotropic.


A convenient choice for the scalar self-interaction potential is
\begin{equation}
	V(|\Phi|) = m^2 |\Phi|^2 + \frac{\kappa}{2} |\Phi|^4,
	\label{eq:potential}
\end{equation}
where $m$ denotes the particle mass and $\kappa$ the self-interaction coupling constant~\cite{Maselli2017}. 
Following~\cite{Colpi1986}, the corresponding equation of state can be written as
\begin{equation}
	p_r = \frac{\rho_0}{3}
	\left(\sqrt{1 + \frac{\rho}{\rho_0}} - 1 \right)^2 ,
	\qquad
	\rho_0 = \frac{m^4}{3\kappa}.
	\label{eq:colpi_eos}
\end{equation}

For strong self-interactions satisfying $\kappa/(4\pi) \gg m^2$, the resulting bosonic configurations are approximately isotropic~\cite{Maselli2017}. 
This formulation reproduces two limiting regimes:
\begin{align}
	p_r &\approx \frac{\rho^2}{12\rho_0},  && (\rho \ll \rho_0), \label{eq:dilute_limit}\\
	p_r &\approx \frac{\rho}{3},            && (\rho \gg \rho_0), \label{eq:ultrarel_limit}
\end{align}
In the limiting case, all models, regardless of the potential, reduce to a polytropic equation of state with index $n=1$ and $\gamma=2$.

Motivated by the dilute-limit correspondence, we approximate the matter content of our dark-matter stars by a simplified polytropic relation:
\begin{equation}
	p_r = K \rho^2, 
	\qquad 
	K = \frac{z}{B},
	\label{eq:polytrope}
\end{equation}
where $z$ is a dimensionless scaling parameter and $B$ a pressure-related constant~\cite{Rindler2014,Hui2017}. 
Unless otherwise specified, the numerical calculations adopt $z = 0.05$ and $B = 205\,\mathrm{MeV/fm^3}$.


Anisotropy naturally arises from gradients in the scalar field. 
Following~\cite{Moraes2021}, it can be expressed as $\sigma = p_t - p_r = -2 e^{-2\lambda}\phi'^2 < 0$. 
Alternatively, Tangphati \textit{et al.}~\cite{Tangphati2025} proposed the phenomenological parametrization
\begin{equation}
	\sigma = \beta\, p_r\, (1 - e^{-2\lambda})=\frac{2 \beta G}{c^{2}r} \, p_{r}(r) m(r),
	\label{eq:anisotropy}
\end{equation}
where the dimensionless coefficient $\beta$ ensures that the anisotropy vanishes at both the stellar center and the surface~\cite{Silva2015,Horvat2011,Cattoen2005,Folomeev2018,Yagi2015}. 
In the weak-field limit ($1 - e^{-\lambda} \ll 1$), $\sigma \approx 0$. 
Since $p_r > p_t$ (as indicated by Eqs.~\ref{eq:pr}$-$\ref{eq:pt}), physical consistency requires $\beta < 0$. 
Accordingly, our numerical analysis considers $\beta$ within the range $[-2, 0]$.

In the following section, these relations are implemented into the modified 4D EGB Tolman$-$Oppenheimer$-$Volkoff equations to analyze the equilibrium configurations and their dependence on the parameters $\alpha$ and $\beta$.



\section{Numerical Analysis of Star Properties}
In this section, an extensive study of the physical properties of anisotropic dark matter stars within the scope of four dimensional Einstein$-$Gauss$-$Bonnet (4D EGB) gravity is documented. Considering the given equation of state ($ p_r = K \rho^2 $) with the anisotropy factor, the interior stellar configurations are acquired by means of the numerical integration of the modified TOV equations as presented in Eqs. (\ref{pr1})-(\ref{m1}). In this manner, we explore, in a numerical context, the manner in which the equilibrium structure and the properties of compact dark matter stars as a whole are modified in light of higher-order curvature corrections and the equilibrium structure anisotropic adjustments.


At the beginning of the numerical integration process, the focus is on the center of the star, where the mass enclosed is zero and pressure is equal to the center pressure $p_{c}$. As for the numerical integration, at the surface of the star ($r=R$), we have the standard boundary conditions $p_{r}(R)=p_{t}(R)=0$ and we take the total mass of the star to be $M=m(r=R)$.

We methodically investigate how the Gauss-Bonnet coupling parameter ($\alpha$) and the anisotropy parameter ($\beta$) influence the internal structure and characteristics of the star. The stability of the produced anisotropic dark matter star configurations integrates various methods and checks the static stability criteria, the sound speed, the adiabatic index, and the energy conditions. Collectively, these criteria serve to comprehensively analyze the equilibrium structure and the dynamical stability of the solutions. In the remaining analysis, we use the stellar radii in kilometers ($km$) and the masses in solar units  ($M_{odot}$).



\subsection{Profiles for variation of rescaled Gauss$-$Bonnet coupling constant}

\begin{figure}[!hbt]
	\centering
	\includegraphics[width=0.5\textwidth]{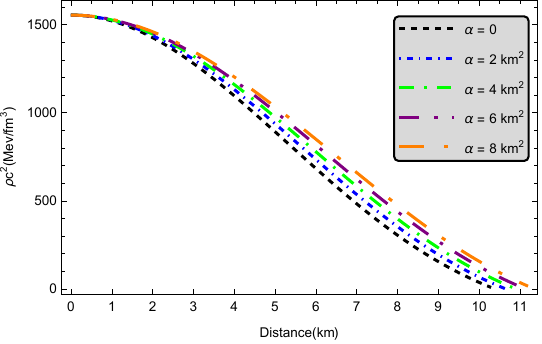}\\[0.5cm]
	\includegraphics[width=0.5\textwidth]{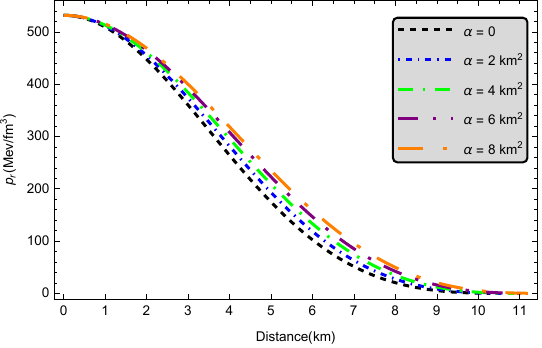}\\[0.5cm]
	\includegraphics[width=0.5\textwidth]{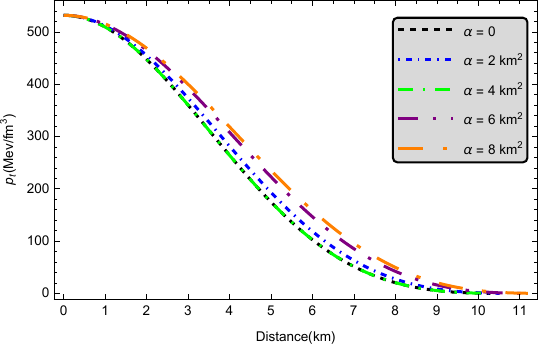}
	\caption{From top to bottom, we present the energy density $\rho$, radial pressure $P_r$, 
		and transverse pressure $P_\perp$ as functions of the radial coordinate $r$. 
		The range of values for $\alpha \in [0,8]\ \mathrm{km}^2$, while the other parameters are fixed as 
		$B = 205~\mathrm{MeV/fm^3}$, $\beta = -0.5$, and $z = 0.05$. 
		A black dashed line represents the anisotropic solution of Einstein's gravity.}
		\label{fig1}
\end{figure}

To examine the influence of the rescaled Gauss$-$Bonnet coupling parameter $\alpha$ on the internal structure of anisotropic dark boson stars, a series of numerical integrations was performed for $\alpha \in [0,8]\ \mathrm{km}^2$, keeping other model parameters fixed ($B = 205\ \mathrm{MeV/fm^3}$, $\beta = -0.5$, and $z = 0.05$). The resulting profiles of the energy density $\rho(r)$, radial pressure $p_r(r)$, and tangential pressure $p_t(r)$ for different values of $\alpha$ are displayed in Fig.~\ref{fig1}. 
\\

\begin{figure}[!hbt]
	\centering
	\includegraphics[width=0.5\textwidth]{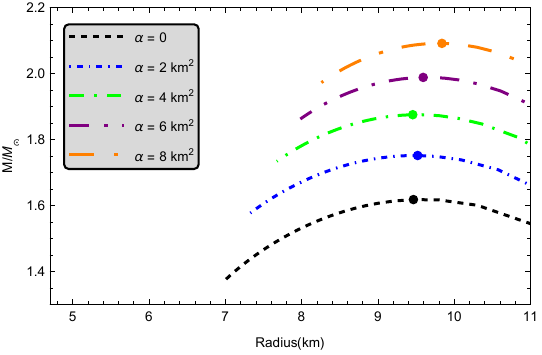}\\[0.5cm]
	\includegraphics[width=0.5\textwidth]{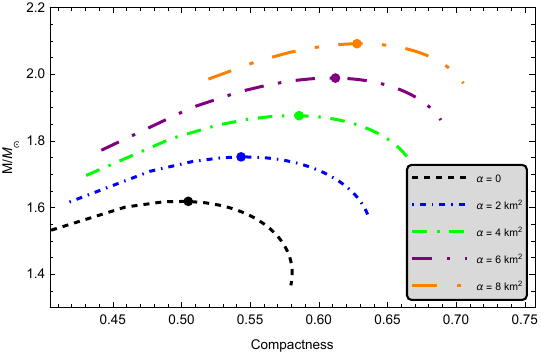}
	\caption{The effect of the coupling constant on the mass-radius relation and the maximum compactness of dark matter compact stars is investigated. These results are obtained using the parameter values employed in Fig. \ref{fig1}. The general relativity (anisotropic solution of Einstein's gravity) result is also indicated by a black dashed line.}
	\label{fig2}
\end{figure}

The computed profiles reveal that increasing $\alpha$ significantly affects both the compactness and mass. As $\alpha$ grows, the maximum mass and corresponding radius of the star increase: the maximum mass rises from
$M_{\mathrm{max}} = 1.61824\,M_\odot \quad (\alpha = 0)$
to $M_{\mathrm{max}} = 2.09145\,M_\odot \quad (\alpha = 8\ \mathrm{km}^2)$. The corresponding radius increases from $R \simeq 9.46712\ \mathrm{km}$ to $R \simeq 9.84084\ \mathrm{km}$. Consequently, the compactness ratio $(2M/R)$ also increases, as summarized in Table~\ref{t1} and illustrated in Fig.~\ref{fig2}.


\renewcommand{\thetable}{\arabic{table}}
\begin{table}
	\caption{Numerical values of $M_{\mathrm{max}}$, $R$, and compactness $(2M/R)$ for representative values of $\alpha$ (fixed $B = 205\ \mathrm{MeV/fm^3}$, $\beta = -0.5$, $z = 0.05$).}
	\label{t1}
	\centering
	\resizebox{\columnwidth}{!}{%
		\begin{tabular}{lrrrr}
			\hline
			$\alpha\ (km^{2})$ & $M[M_{\odot}]$ & $R_{M}[km]$ & $p_{c}[Mev/fm^{3}] $& $2M/R$  \\
			\hline
			$0.0$ & $1.61824$ & $9.46712$  & $1124$&$0.50494$ \quad\\
			$2.0$ & $1.75179$ & $9.52284$ & $1404 $&$0.54341$ \quad  \\
			$4.0$ & $1.87537$ & $9.45858$ & $1966$&$0.58570$ \quad   \\
			$6.0$ & $1.98834$ & $9.59507$  & $2248$&$0.61215$ \quad  \\
			$8.0$ & $2.09145$ & $9.84084$  & $2248$&$0.62781$ \quad  \\
			\hline
		\end{tabular}%
	}
\end{table}

As $\alpha$ increases, the star compactness $2M/R$ increases too, although it stays in the permissible zone and upholds the modified Buchdahl inequality \cite{Chakraborty2020,Matthew2015} Given by:
\begin{equation}
	\frac{2M}{R} < \frac{8}{9}\left(1+\frac{\alpha}{R^2}\right).
\end{equation}
Consequently, the Gauss$-$Bonnet contribution allows for increasingly compact configurations that do not form apparent horizons, keeping the bosonic stars outside the black hole regime. The results show that the coupling parameter $\alpha$ has considerable influence on the mass and radius of the star while keeping the star in the state of static stability.

\subsection{Profiles for variation of the anisotropy parameter}
\begin{figure}[!hbt]
	\centering
	\includegraphics[width=0.5\textwidth]{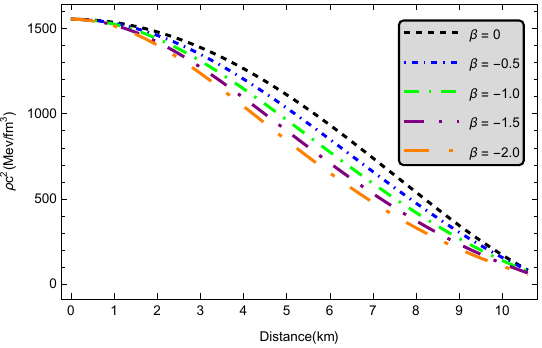}\\[0.5cm]
	\includegraphics[width=0.5\textwidth]{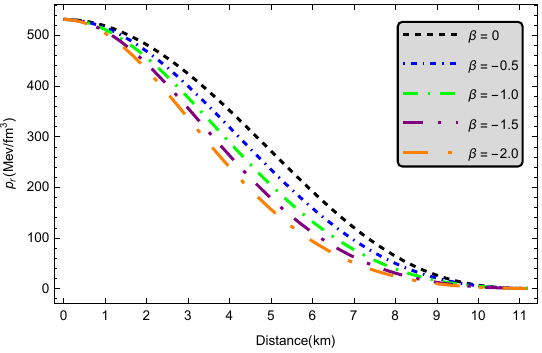}\\[0.5cm]
	\includegraphics[width=0.5\textwidth]{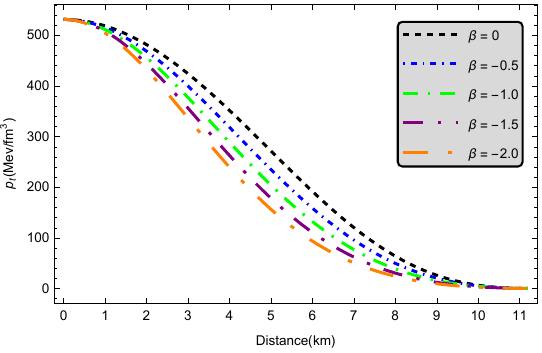}
	\caption{From top to bottom, we present the energy  density $\rho$, radial pressure $p_r$, 
		and transverse pressure $p_t$ as functions of the radial coordinate $r$. 
		The range of values for $\beta \in [-2,0]$, while the other parameters are fixed as 
		$B = 205~\mathrm{MeV/fm^3}$, $\alpha=8km^2$, and $z = 0.05$. 	A black dashed line represents the isotropic solution of 4D EGB gravity (weak-field limit).}
	\label{fig3}
\end{figure}

The impact of the anisotropy parameter $\beta$ on the energy density, radial pressure ($p_r$), and tangential pressure ($p_t$) in the structure of dark matter stars is illustrated in Figure $\ref{fig3}$, alongside the isotropic case of the Gauss-Bonnet gravity for comparison. The parameter sets represent different values of $\beta$ in the range $\beta \in [-2.0,0.0]$, with the corresponding values $B=205~\mathrm{MeV/fm^3}$, $z = 0.05$, and $\alpha= 8km^2$.


The relations $(M-R)$ and $(M-2M/R)$ are shown in Fig. ref{fig4}. Since we are talking about dark matter stars, we take a range of negative $\beta$ values; refer to Sect. 2 for more. $\beta$ significantly affects the ($M-R$) relations, as its greater values directly correlate to an increase in maximum mass. For the anisotropic case, the maximum mass, $M_{max} = 1.97M_{\odot}$ ($\beta = -0.5$), occurs at the upper value in the range provided in Table $\ref{t2}$, and the maximum mass radius is $R_{max} = 11.30 km$. The maximum mass of dark matter stars for $\beta = 0$ (the isotropic solution in four-dimensional Gauss$-$Bonnet gravity) is $2.14M_{\odot}$.

\begin{figure}[!hbt]
	\centering
	\includegraphics[width=0.5\textwidth]{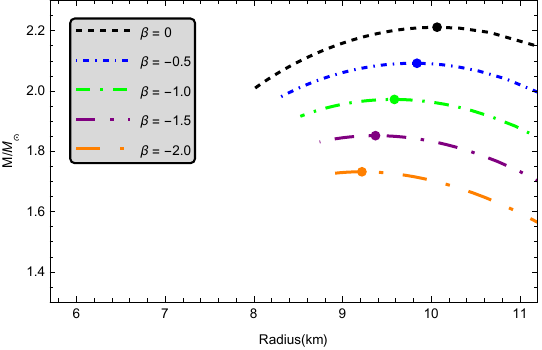}\\[0.5cm]
	\includegraphics[width=0.5\textwidth]{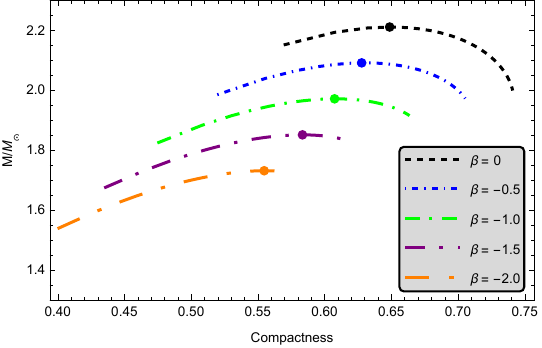}
	\caption{The effect of the anisotropy parameter $\beta$ on the mass-radius and maximum compactness relations of dark matter compact stars is examined. The results are obtained using the parameter values presented in Table \ref{t2}. The isotropic case (4D EGB gravity) is illustrated by a black dashed line.}
	\label{fig4}
\end{figure}

Analysis results of the impact of dark matter in the proposed stellar models indicate that under negative $\beta$ conditions, isotropic solutions in 4D EGB gravity correspond to higher masses in comparison to their anisotropic counterparts (see Fig.  \ref{fig4}). In the lower section of Figure $\ref{fig4}$, the influence of the beta parameter on the characteristics of the ($M-2M/R$) relation is presented. As indicated in Table $\ref{t2}$, the star's maximum compactness increases with beta, achieving $2M/R \approx 0.628$ at $\beta= -0.5$. 

\begin{table}
	\caption{A summary of the structural properties of dark matter stars is presented for 
		$B=205~\mathrm{MeV/fm^3}$, 
		$z=0.05$, $\alpha = 8\ km^2$, and varying values of $\beta$.}
	\label{t2}
	\centering
	\resizebox{\columnwidth}{!}{%
		\begin{tabular}{lrrrr}
			\hline
			$\beta$ & $M[M_{\odot}]$ & $R_{M}[km]$ & $p_{c}[Mev/fm^{3}] $& $2M/R$    \\
			\hline
			$0.0$ & $2.21106$ & $10.0673$  & $1404$&$0.648783$ \quad\\
			$-0.5$ & $2.09145$ & $9.84084$ & $2247 $&$0.627808$ \quad  \\
			$-1.0$ & $1.97149$ & $9.58647$ & $3933$&$0.607501$ \quad   \\
			$-1.5$ & $1.85129$ & $9.37330$  & $7020$&$0.583437$ \quad  \\
			$-2.0$ & $1.73151$ & $9.22028$  & $12360$&$0.554744$ \quad  \\
			\hline
		\end{tabular}%
	}
\end{table}

\section{stability, Causality and energy conditions}
In this section, the stability of the proposed model is assessed to ensure that the structure under consideration remains stable under varying conditions. For this purpose, the static stability criterion, the adiabatic index, and the speed of sound are employed; each of these criteria is systematically analyzed, and the resulting data are presented graphically to clearly illustrate the variations and stability behavior of the model.

\subsection{Static stability criterion in terms of central pressure}
In this work, the stability of equilibrium configurations of bosonic stars is examined within the framework of four-dimensional Einstein$-$Gauss$-$Bonnet ($4D EGB$) gravity, with a particular focus on the static stability criterion (see Ref. \cite{Harrison1965,Zeldovich1971}, for more details). The results are displayed in the $M-p_{c}$ plane, where $M$ denotes the gravitational mass and $p_{c}$ represents the central pressure. Although this criterion thoroughly investigated within General Relativity (GR), its relevance in modified gravity scenarios has also been emphasized in Refs. \cite{Tangphati2021ann,Maulana2019,Sham2012,Pretel2022}.The criterion may be written as:
\begin{equation}
	\frac{d M}{d \rho_c} > 0 \to \text{stable configuration},
\end{equation}
\begin{equation}
	\frac{d M}{d \rho_c} < 0 \to \text{unstable configuration}.
\end{equation}

Because the equation of state (EoS) provides a monotonic relation between pressure and energy density, one may write
\begin{equation}
	\frac{dM}{dp_{c}} \;=\; \frac{dM}{d\rho_{c}}\,\frac{d\rho_{c}}{dp_{c}}.
\end{equation}
for physically causal EoSs, since $d\rho_{c}/dp_{c}>0$, $dM/dp_{c}$ has the same sign as $dM/d\rho_{c}$. Thus, the stability criterion stated in terms of $\rho_{c}$ is equivalent to the criterion stated in terms of the central pressure $p_{c}$.

\begin{figure}[!hbt]
	\centering
	\includegraphics[width=0.5\textwidth]{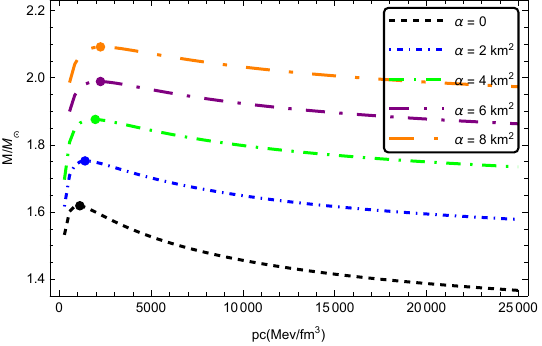}\\[0.5cm]
	\includegraphics[width=0.5\textwidth]{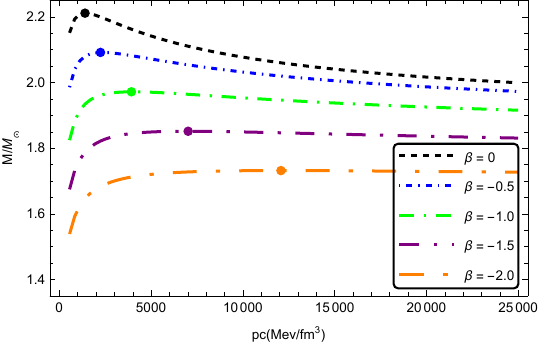}
	\caption{The $M-p_{c}$ curves for a family of anisotropic dark matter stars with variations in $\alpha$ and $\beta$ are presented.}
	\label{fig5}
\end{figure}

For each case considered for anisotropic dark-matter stars, with variations in the Gauss$-$Bonnet coupling $\alpha$ and the anisotropy parameter $\beta$, figure~\ref{fig5} presents the $M$--$p_{c}$ relations. The different curves correspond to distinct values of $\alpha$ and $\beta$. Having a larger $\alpha$ value shifts the curves up, resulting in configurations with larger maximum masses, while negative values of $\beta$ tend to shift the $M(p_{c})$ curve toward configurations with smaller maximum masses. The transition from stable to unstable one is also illustrated in Figure~\ref{fig5}. The transition occurs at $dM/dp_{c} = 0$ with $dM/dp_{c} > 0$ for stable configurations. This condition is necessary for stability but not sufficient.




\subsection{Sound speed and causality, and Adiabatic indices}
The sound speeds in the radial and tangential directions are defined, respectively, as follows:
\begin{equation}
	v_r^2 = \frac{d p_r}{d \rho}, \quad v_t^2 = \frac{d p_t}{d \rho}.
\end{equation}

To ensure causality, these quantities must satisfy
\begin{equation}
	0 < v_{r,t}^2 < c^2,
\end{equation}
\begin{figure}[!hbt]
	\centering
	\includegraphics[width=0.5\textwidth]{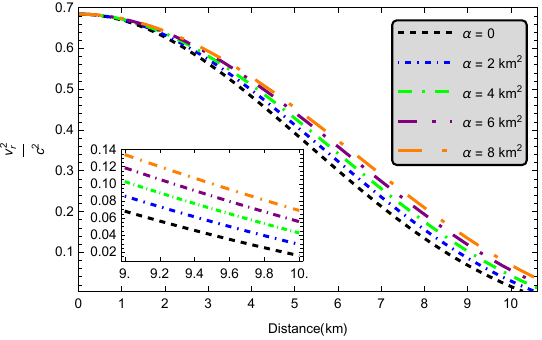}\\[0.5cm]
	\includegraphics[width=0.5\textwidth]{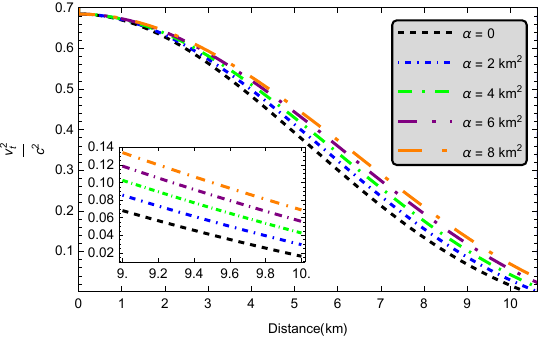}
	\caption{The squared sound speed in the radial and tangential directions for anisotropic dark matter compact stars with variations in the parameter $\alpha$ is investigated.}
	\label{fig6}
\end{figure}
\begin{figure}[!hbt]
	\centering
	\includegraphics[width=0.5\textwidth]{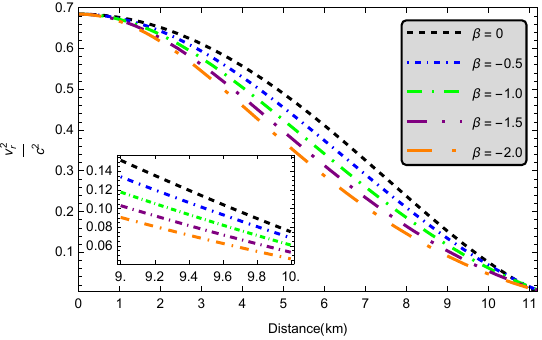}\\[0.5cm]
	\includegraphics[width=0.5\textwidth]{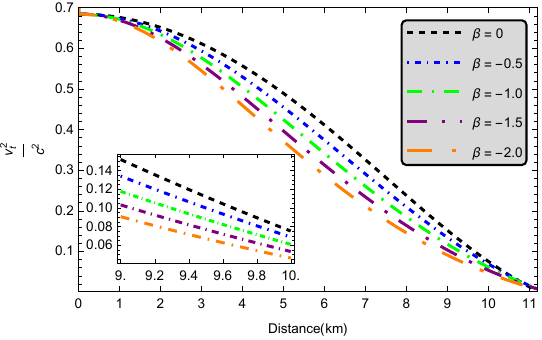}
	\caption{The squared sound speed in the radial and tangential directions for anisotropic dark matter stars with variations in $\beta$.}
	\label{fig7}
\end{figure}
throughout the anisotropic dark matter boson star. 
Figures~\ref{fig6} and \ref{fig7} show the radial and tangential sound speeds for representative values of the model parameters $\alpha$ and $\beta$. The results indicate that the sound speeds remain strictly within the physically allowed range, confirming that causality is preserved.

The dynamical stability of the anisotropic dark matter boson star is investigated through the radial adiabatic index $\Gamma_r$, originally introduced by Chandrasekhar~\cite{Chandrasekhar1964} and further generalized for anisotropic configurations in four-dimensional Einstein$-$Gauss$-$Bonnet gravity~\cite{Banerjee2024EPJC}. For radial perturbations, $\Gamma_{r}$ is given by
\begin{equation}
	\Gamma_{r} = v_r^2 \left( 1 + \frac{\rho}{p_r} \right).
\end{equation}

\begin{figure}[!hbt]
	\centering
	\includegraphics[width=0.5\textwidth]{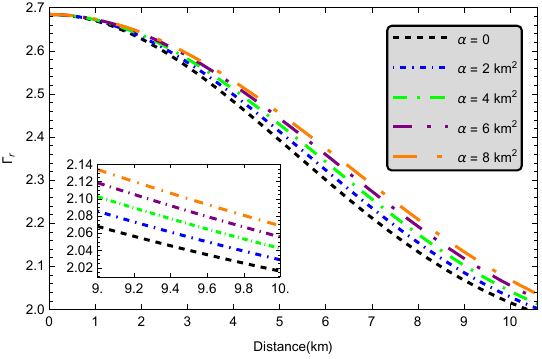}\\[0.5cm]
	\includegraphics[width=0.5\textwidth]{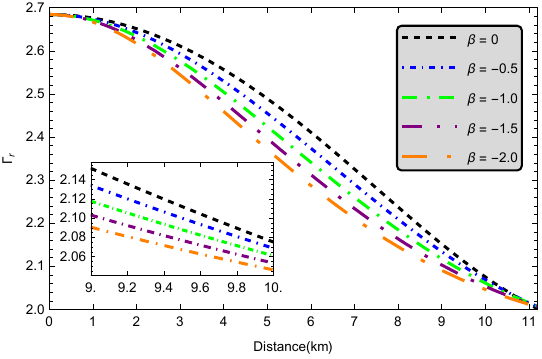}
	\caption{Radial profiles of the adiabatic index $\Gamma_{r}$ for anisotropic dark matter stars with varying $\alpha$ and $\beta$.}
	\label{fig8}
\end{figure}

A configuration is considered dynamically stable if 
$\Gamma_{r}>\Gamma_{cr}$, where 
$\Gamma_{cr}=4/3$ corresponds to the critical adiabatic index for homogeneous isotropic spheres~\cite{Glass1983}. Figure~\ref{fig8} presents the radial dependence of 
$\Gamma_{r}$
for several values of 
$\alpha$ and 
$\beta$. Results indicate that $\Gamma_{r}$ declines steadily with increasing $r$ while remaining above $\Gamma_{cr}$ which indicates a dynamically stable configuration.

\subsection{Energy Conditions}

Finally, the obtained solutions must be capable of describing realistic astrophysical configurations. Therefore, as an additional consistency test, we examine whether the energy conditions are satisfied.
To that end, the conditions \cite{Mak2002,Deb2018,Deb2017,Bhar2017,Panotopoulos2019}
\begin{equation}
	\rho \geq 0\,,
	\label{null}
\end{equation}
\begin{equation}
	\rho + p_{r,t}  \geq  0\,,
\end{equation}
\begin{equation}
	\rho - p_{r,t}  \geq  0\,,
\end{equation}
\begin{equation}
	E_+ \equiv \rho + p_r + 2 p_t \geq 0\,,
\end{equation}
\begin{equation}
	E_- \equiv \rho - p_r - 2 p_t \geq 0\,,
	\label{e-}
\end{equation}
are investigated.

\begin{figure*}[!hbt]
		{\includegraphics[width=.45\linewidth]
			{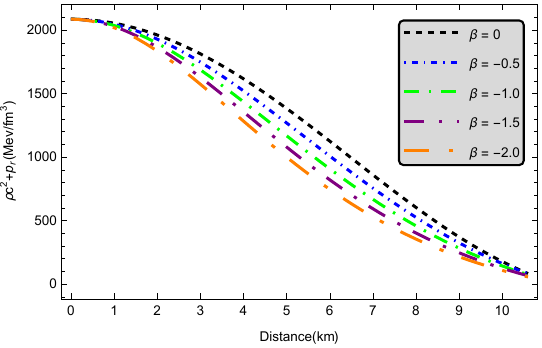}
			\includegraphics[width=.45\linewidth]
			{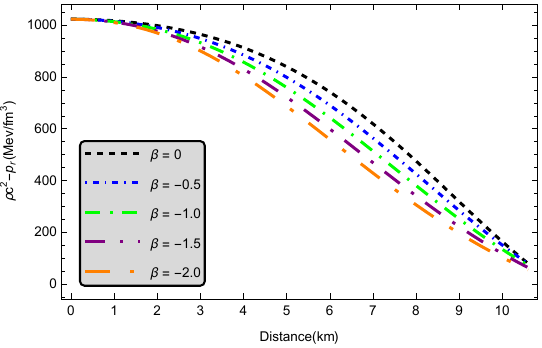}
			\includegraphics[width=.45\linewidth]
			{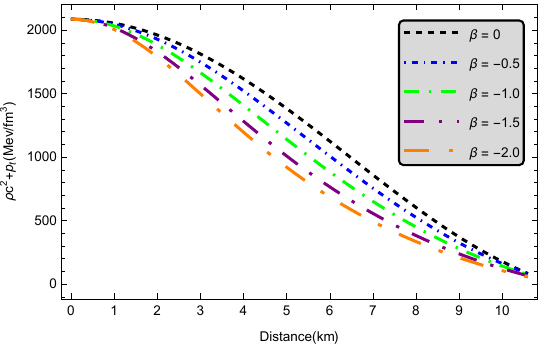}
			\includegraphics[width=.45\linewidth]
			{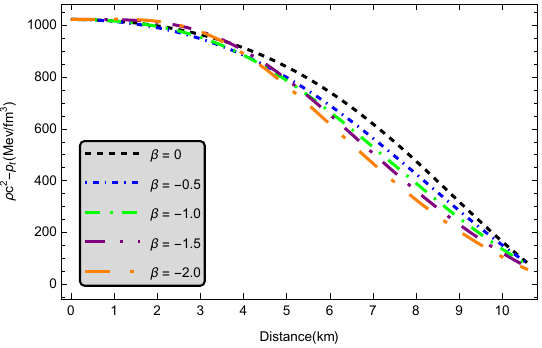}
			\includegraphics[width=.45\linewidth]
			{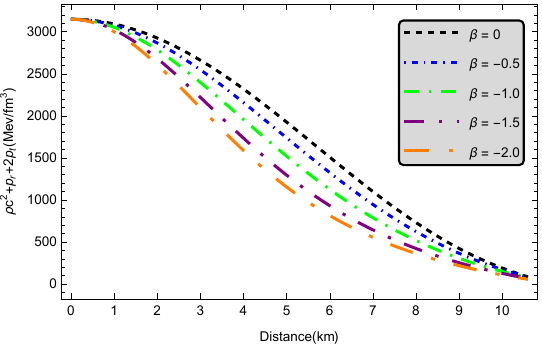}
			\includegraphics[width=.45\linewidth]
			{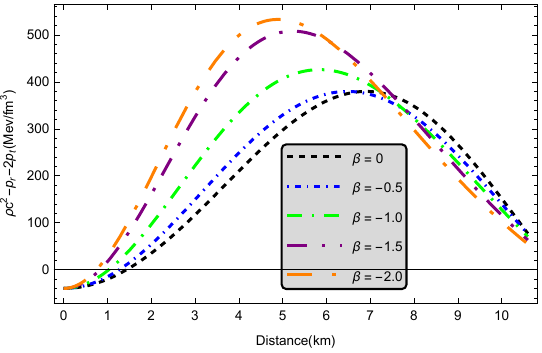}
		}
		\caption{{\protect\small 
				Profiles of the Energy Conditions for anisotropic dark matter stars with varying $\beta$.
		}}
		\label{EC}
\end{figure*}
The null energy condition (\ref{null}), as illustrated in the first panel of Figure \ref{fig3}, together with the other energy conditions, except for the last one in Eq. (\ref{e-}), are satisfied within the 4D EGB gravity framework for all considered values of the anisotropy parameter $\beta$ (see Figures \ref{EC}). The final energy condition becomes fully satisfied for $\beta \in [-2, 0]$ at radial distances greater than approximately $0.8-1.5\ km$ (see last panel of Figure \ref{EC}).
This confirms that the matter distribution remains physically acceptable.

\section{Concluding remarks}
In this study, we investigated the structure and stability of anisotropic bosonic dark matter stars within the framework of the regularized four-dimensional Einstein--Gauss--Bonnet (4D EGB) gravity. By numerically integrating the modified Tolman--Oppenheimer--Volkoff equations for an anisotropic polytropic equation of state, we analyzed the effects of the Gauss--Bonnet coupling parameter ($\alpha$) and the anisotropy parameter ($\beta$) on the equilibrium and stability of dark matter configurations.

Our results show that increasing the Gauss--Bonnet coupling enhances both the maximum mass and compactness of the dark matter stars---from about $M_{\text{max}} \approx 1.62\,M_\odot$ at $\alpha = 0$ to $\approx 2.09\,M_\odot$ at $\alpha = 8\,\text{km}^2$. In contrast, negative anisotropy ($\beta < 0$) reduces these quantities, lowering the maximum mass to $\approx 1.73\,M_\odot$ at $\beta = -2$. All configurations remain statically stable up to the maximum-mass point and satisfy causality as well as all relevant energy conditions.

These findings indicate that the interplay between anisotropy and higher-curvature corrections in 4D EGB gravity plays a crucial role in determining the structure and physical limits of compact dark matter stars. The model provides an extended theoretical framework that can potentially be linked to observable features of dark compact objects, such as gravitational lensing or gravitational-wave signatures.

Future work could focus on extending this analysis to include rotating or magnetized configurations, as well as exploring other modified gravity theories such as $F(R, T)$ or scalar--tensor extensions, to further constrain the nature of self-gravitating dark matter systems.




\section*{Acknowledgments}
I thank Research Council of Shiraz University and Persian Gulf University for their support.


\end{document}